\documentstyle [12pt]{article}

\begin{document}

\begin{center}
\noindent{\bf EFFECTIVE LAGRANGIAN FOR BARYONS AND BARYON-MESON INTERACTION}
\end{center}

\begin{center}
{\sf S.S. Olean}
\end{center}

\begin{center}
{\sf Abstract}
\end{center}

{\sf Expansion of model Nambu-Jona-Lasinio is considered and on its basis an 
effective lagrangian for mesons, baryons and baryon - meson interaction has 
been obtained.}

\newpage 
{\sf The successes of quantum chromodynamics (QCD) in the description of 
high-energy interactions of hadrons are well- known. However its use has met 
many difficulties in medium and especially in low-energy range. The main 
difficulty is that the chromodynamical coupling constant }$\alpha _c ${\sf 
in this range becomes large, that is why the common perturbation theory 
becomes inapplicable. Therefore the description of low-energy processes and 
properties of elementary particles requires putting forward the other 
approximate methods and models. Some of them are related to QCD. In this 
article the method allowing to derive the reported phenomenological meson 
and baryon Lagrangians, successfully describing low-energy physics of 
hadrons on the basis of effective quark interaction has been given.}

{\sf Here we shall not discuss the constructing of an effective quark 
Lagrangian on the basis of QCD Lagrangian being examined for 4-th quark 
Lagrangian in [1,2].} {\sf We shall consider effective quark Lagrangian that 
is an extension of Lagrangian offered in [3,4]:}

\begin{equation}
\label{eq1}
\begin{array}{l}
 L = \overline q (i\gamma _\mu \partial _\mu - m_q )q + \frac{G_1 
}{2}[(\overline q \lambda ^aq)^2 + (\overline q i\gamma _5 \lambda ^aq)^2] - 
\\ 
 \frac{G_2 }{2}[(\overline q \gamma _\mu \lambda ^aq)^2 + (\overline q 
\gamma _5 \gamma _\mu \lambda ^aq)^2] + \sum {C_i (\overline q \Gamma _i q)} 
^4, \\ 
 \end{array}
\end{equation}

{\sf where }$\Gamma _i ${\sf -matrices, i=($\mu $,a,$\alpha )$, $\mu $- 
spatial index, a- flavour, $\alpha $- colour. We shall now show, how meson 
and baryon fields are introduced and how phenomenological meson and baryon 
Lagrangians are obtained. With generating functional }$W(\overline \eta 
,\eta )${\sf this procedure can be conducted in three steps:}

\begin{equation}
\label{eq2}
W(\overline \eta ,\eta ) = \frac{1}{N}\int {dqd\overline q } \exp 
[L(q,\overline q ) + \eta \overline q + \overline \eta q],
\end{equation}

{\sf where}

\[
\begin{array}{l}
 L(q,\overline q ) = \overline q (i\gamma _\mu \partial _\mu - m_q^0 )q + 
\frac{G_1 }{2}[(\overline q \lambda ^aq)^2 + (\overline q i\gamma _5 \lambda 
^aq)^2] - \\ 
 \frac{G_2 }{2}[(\overline q \gamma _\mu \lambda ^aq)^2 + (\overline q 
\gamma _\mu \gamma _5 \lambda ^aq)^2] + \\ 
 \sum\limits_{i,I} {\frac{\sqrt 2 G_i C^2}{2}[g_I (\overline q } \overline q 
\overline q T)_k^j (\lambda ^a)_j^i \Gamma _I (Tqqq)_i^k \overline q _s 
(\lambda ^a)_t^s \Gamma _I q^t + \\ 
 h_I (\overline q \overline q \overline q T)_k^j \Gamma _I (Tqqq)_j^i 
(\lambda ^a)_i^k \overline q _s (\lambda ^a)_t^s \Gamma _I q^t]. \\ 
 \end{array}
\]

{\sf In formula (\ref{eq2}) I= S, P, V, A, and }$\Gamma _I ${\sf - correspond to 
}$I,i\gamma _5 ,\gamma _\mu ,\gamma _\mu \gamma _5 ${\sf , }$G_i ${\sf - is 
equal to }$G_1 ${\sf for S- and P- variants and to }$G_2 ${\sf for V- and A- 
variants,}

\[
(Tqqq)_i^k = (R_{Vi;i_1 i_2 i_3 }^k + \mathop R\nolimits_{Vi;i_1 i_2 i_3 }^k 
)q_{b_1 }^{i_1 } q_{b_2 }^{i_2 } q_{b_3 }^{i_3 } \varepsilon ^{b_1 b_2 b_3 
},
\]

{\sf and the explicit forms for }$R_T ${\sf and }$R_V ${\sf are given in 
[5].We shall proceed from the Lagrangian (\ref{eq2}) to Lagrangian containing meson 
and baryon fields in addition to quark fields:}

\begin{equation}
\label{eq3}
\begin{array}{l}
 {L}' = \overline q [i\gamma _\mu \partial _\mu - m_0 + S + i\gamma _5 \Pi + 
\gamma _\mu V_\mu + \gamma _5 \gamma { }_\mu A_\mu + \\ 
 \sqrt 2 G_1 g_1 \overline B \lambda ^aB\lambda _a - \sqrt 2 G_1 h_1 
\overline B B\lambda ^a\lambda _a + \sqrt 2 G_1 g_2 \overline B \gamma _5 
\lambda ^aB\gamma _5 \lambda _a - \\ 
 \sqrt 2 G_1 h_2 \overline B \gamma _5 B\lambda ^a\gamma ^5\lambda _a + 
\sqrt 2 G_2 g_3 \overline B \gamma _\mu \lambda ^aB\gamma _\mu \lambda _a - 
\\ 
 \sqrt 2 G_2 h_3 \overline B \gamma _\mu B\lambda ^a\gamma _\mu \lambda _a + 
\sqrt 2 G_2 g_4 \overline B \gamma _5 \gamma _\mu \lambda ^aB\gamma _5 
\gamma _\mu \lambda _a - \\ 
 \sqrt 2 G_2 h_4 \overline B \gamma _5 \gamma _\mu B\lambda ^a\gamma _5 
\gamma _\mu \lambda _a ]q - \\ 
 \frac{S_a^2 + \Pi _a^2 }{2G_1 } - \frac{V_a^{\mu 2} + A_a^{\mu 2} }{2G_2 } 
+ \frac{g_1 }{\sqrt 2 }\overline B SB - \frac{h_1 }{\sqrt 2 }\overline B BS 
+ \frac{g_2 }{\sqrt 2 }\overline B \gamma _5 \Pi B - \\ 
 \frac{h_2 }{\sqrt 2 }\overline B \gamma _5 B\Pi + \frac{g_3 }{\sqrt 2 
}\overline B \gamma _\mu V_\mu B - \frac{h_3 }{\sqrt 2 }\overline B \gamma 
_\mu BV_\mu + \frac{g_4 }{\sqrt 2 }\overline B \gamma _\mu \gamma _5 A_\mu B 
- \\ 
 \frac{h_4 }{\sqrt 2 }\overline B \gamma _\mu \gamma _5 BA_\mu , \\ 
 \end{array}
\end{equation}

{\sf where B- octet of baryons , }$S,\Pi ,V_\mu ,A_\mu - ${\sf scalar, 
pseudo-scalar, vector and axial mesons respectively.}

{\sf Lagrangian for standard $\sigma $-model, which describes scalar and 
pseudo-scalar mesons interaction has been obtained from Lagrangian (\ref{eq3}) in 
[4]. In addition to Lagrangian for this model we shall obtain an effective 
Lagrangian for baryons. For that we shall consider the divergent quark loops 
of four types [4] and the diagrams of the mass operator type for baryons. In 
these loops there are different divergencies for mesons and baryons. The sum 
of the meson diagrams with quark loops results in the expression:}

\[
\begin{array}{l}
 Tr\left\{ {\left[ {p^2I_2 + 2\left( {I_1 + M^2I_2 } \right)} \right]} 
\right.\times \left[ {\left( {s - M} \right)^2 + \Pi ^2} \right] - \\ 
 I_2 \left( {\left[ {\left( {s - M} \right)^2 + \Pi ^2} \right]^2 - \left[ 
{\left( {s - M} \right),\Pi } \right]_ - ^2 } \right) \\ 
 \end{array}
\]

{\sf Here the first term is a kinetic term (p-momentum of a meson). This 
term completely defines renormalization of meson fields. }$I_1 ${\sf and 
}$I_2 ${\sf - divergent integrals, which explicit forms have been given in 
[3], and }$S = gS^R,\Pi = g\Pi ^R${\sf , where }$g = \left( {4I_2 } 
\right)^{ - \frac{1}{2}}${\sf . Performing similar computations for baryons 
we obtain from Lagrangian (\ref{eq3}): }

\[
\overline u (p)\left\{ { - i\gamma _\mu p_\mu I_3 + I_4 } \right\}u(p),
\]

{\sf where}

\[
\begin{array}{l}
 I_{3,4} = \frac{3i}{2(2\pi )^4V}[A^2K_{3,4} (m_p ,m_u ,m_u ) + B^2K_{3,4} 
(m_p ,m_d ,m_d ) + \\ 
 C^2K_{3,4} (m_p ,m_s ,m_s ) + D^2K_{3,4} (m_n ,m_u ,m_d ) + E^2K_{3,4} 
(m_\Lambda ,m_u ,m_s ) + \\ 
 F^2K_{3,4} (m_\Sigma ,m_u ,m_s ) + G^2K_{3,4} (m_\Sigma ,m_d ,m_s )], \\ 
 \end{array}
\]

\[
\begin{array}{l}
 K_3 (m_M ,m_1 ,m_2 ) = \int {\frac{d^4sd^4t8(i\gamma _\mu s_\mu - m_2 
)(2i\gamma _\nu t_\nu + m_1 )}{[(p + t + s)^2 + M^2](s^2 + m_2^2 )(t^2 + 
m_1^2 )}} , \\ 
 K_4 (m_M ,m_1 ,m_2 ) = \int {\frac{d^4sd^4t4[2i\gamma _\mu (t_\mu - s_\mu ) 
+ M]( - i\gamma _\nu s_\nu + m_2 )(2i\gamma _\alpha t_\alpha + m_1 )}{[(p + 
t + s)^2 + M^2](s^2 + m{ }_2^2 )(t^2 + m_1^2 )}} . \\ 
 \end{array}
\]

{\sf After going to the other fields }$\bar {P}(p) = \sqrt {I_3 } \bar 
{u}(p),P(p) = \sqrt {I_3 } u(p)${\sf we have:}

\[
L = \overline P (p)\left\{ { - i\gamma _\mu p_\mu + \frac{I_4 }{I_3 }} 
\right\}P(p).
\]

{\sf Thus an effective Lagrangian for protons has been obtained. An 
effective Lagrangian for other baryons can be derived in the same manner.}

\begin{center}
{\sf REFERENCES}
\end{center}

{\sf 1.N.I.Karchev and A.A.Slavnov, Teor.Mat.Fiz. (USSR) 65 (1985) p. 
192-200.}

{\sf 2. Goldman T. and Haymaker R.W.-Phys.Rev. (1981), v. 24D, p. 724-751.}

{\sf 3. Nambu I. and Jona-Lasinio G.-Phys.Rev. 122 (1961) 345.}

{\sf Nambu I. and Jona-Lasinio G.-Phys.Rev. 124 (1961) 246.}

{\sf 4. Volkov M. K.-Sov.J.Part.and Nuclei, (1986),v.17,p.186- 203.}

{\sf Volkov M. K.-Sov.J.Part.and Nuclei, (1993),v24,p.81-139.}

5. Efimov G.V., Ivanov M. A. and Lyubovitskij V.E.-JINR Preprint E2-90-24, 
Dubna (1990).

\end{document}